\documentclass[%
aps,
prb,
twocolumn,
superscriptaddress,
floatfix,
notitlepage,
a4paper,
10pt
]{revtex4-2}

\usepackage[english]{babel}
\usepackage[utf8]{inputenc}
\usepackage{amssymb}
\usepackage{amsmath}
\usepackage{xcolor}
\usepackage{graphicx}
\usepackage{pdfsync}
\usepackage{hyperref}
\hypersetup{colorlinks, linkcolor={red}, citecolor={blue}, urlcolor={blue}}
\usepackage{bm}
\usepackage{comment}
\usepackage{mathptmx}

\usepackage{todonotes}

\usepackage[normalem]{ulem}
\usepackage[caption=false]{subfig}

\begin{document}
	
	\title{Statistical Insight into the Correlation of Geometry and Spectral Emission in Network Lasers}
	
	\author{Camillo Tassi}
	\email{camillo.tassi@gmail.com}
	\affiliation{Dipartimento di Fisica, Università di Pisa, Largo Bruno Pontecorvo 3, I-56127 Pisa, Italy}
	
	\author{Riccardo Mannella}
	\affiliation{Dipartimento di Fisica, Università di Pisa, Largo Bruno Pontecorvo 3, I-56127 Pisa, Italy}
	
	\author{Andrea Tomadin}
	\affiliation{Dipartimento di Fisica, Università di Pisa, Largo Bruno Pontecorvo 3, I-56127 Pisa, Italy}
	
	\author{Andrea Camposeo}
	\affiliation{NEST, Istituto Nanoscienze - CNR and Scuola Normale Superiore, Piazza San Silvestro 12, I-56127 Pisa, Italy}
	
	\author{Dario Pisignano}
	\affiliation{Dipartimento di Fisica, Università di Pisa, Largo Bruno Pontecorvo 3, I-56127 Pisa, Italy}
	\affiliation{NEST, Istituto Nanoscienze - CNR and Scuola Normale Superiore, Piazza San Silvestro 12, I-56127 Pisa, Italy}
	
	%\date{}

\begin{abstract}

   Optically active networks show feature-rich emission that depends on the fine details of their geometry, and find diverse applications in random lasers, sensing devices and photonics processors. In these and other systems, a thorough and predictive characterization of how the network geometry correlates with the resulting emission spectrum would be highly important, however such outright description is still lacking.
    In this work, we take a step toward filling this gap, by using the well-known Steady--State ab Initio Laser Theory equations to carry out an extensive set of statistical analyses and establish connections between the random network geometry and their ultimate emission spectrum. Our results show that edge crowding (abundance of short edges in the network) is key to tune the uniformity of the modal intensity distribution of the emission spectrum. A statistical framework for the comprehensive understanding of the network statistical properties is highly significant to establish precise design rules for network-based photonic devices and intelligent systems.

\end{abstract}

\maketitle

\textit{Introduction.}---Network lasers (NLs) have emerged as innovative random laser sources with highly versatile emission properties~\cite{lepri2017complex,gaio2019nanophotonic,giacomelli2019optical,saxena2022sensitivity,consoli2023networks}.   In these systems, the gain medium is made of interconnected, active waveguides as edges of a disordered optical network. Such network supports light amplification along the interconnected edges, while multiple scattering at the network nodes and interference across links generate copious optical modes. The spectral features and intensities of the resulting lasing peaks are determined by the specific network architecture. Lasers can be realized by interconnecting single-mode optical fibers with commercially available components (such as splitters, circulators, and isolators)~\cite{lepri2017complex,giacomelli2019optical}, organo-lead halide perovskites \cite{dhanker2014}, amyloid fibrils conjugated with fluorophores \cite{gong2021}, lithographically designed InP features \cite{saxena2025designed}, or polymer nanofibers electrospun upon extrusion of organic solutions under high voltage~\cite{gaio2019nanophotonic}. 
NLs have high potential for biosensing applications, due to their high sensitivity to changes in the local environment~\cite{gaio2019nanophotonic}.
Moreover, their quasi-2D geometry and material compatibility make them promising for on-chip integration~\cite{saxena2025designed}, with potential use in programmable light sources~\cite{saxena2022sensitivity}, and neuromorphic optical processors~\cite{shen2017deep,ng2024retinomorphic}.

Due to their disordered geometry, a full description of NLs presents several issues. In fact, a key challenge lies in establishing a robust correlation between the network geometry and the characteristics of emission.  Simulations ~\cite{gaio2019nanophotonic} have shown a decrease of the lasing threshold upon increasing the network degree, i.e., the average number of connections per node.  Another study highlighted the high sensitivity of the NL spectrum to the spatial shape of the pump profile~\cite{saxena2022sensitivity}. 
In interference-based interpretations of random lasing, multiple scattering has been proposed to form loop optical paths acting as effective  cavities~\cite{cao1998ultraviolet}, while the optical response of complex structures, including networks, can articulate in modes at very various degree of localization~\cite{wiersma2016clear}; in network lasers, the statistics of paths and geometries involved in lasing is prevalently controlled by the distribution of edge lengths.

Given the subtle dependence of NL emission on the underlying geometrical configuration, a statistical framework offers the most appropriate means to analyze and interpret the spectra. Statistical studies in this context generally rely on a structural, discrete characterization of the network, such as its degree~\cite{gaio2019nanophotonic}. In this work, instead, we characterize a statistical sample of nanofiber-based NL using a spectral inverse participation ratio (SIPR) as continuous indicator, which reflects the hypothesis that the length distribution of fiber segments between pairs of nodes (i.e. the edges of the network) plays a crucial role in shaping the modal intensity distribution. This approach enables a more quantitative investigation of the correlations between the geometry of the network and its optical properties. To study the emission spectrum, we use the Steady-state Ab Initio Laser Theory (SALT), which allows one to compute the lasing modes and thresholds by solving a non-Hermitian eigenvalue problem, derived from Maxwell’s equations coupled to the gain medium. This theory assumes stationary inversion of the gain medium, which significantly simplifies the description of NLs~\cite{tureci2006self,tureci2008ab,ge2010steady_th,ge2010steady}.  In order to solve the set of nonlinear equations that determine the spectrum for a statistically significant sample of networks, we employ the single-pole approximation (SPA-SALT)~\cite{ge2010steady_th,ge2010steady,Julia,bezanson2017julia}. The passive cavity modes are obtained by numerically solving a nonlinear eigenvalue problem, using the Argument Principle followed by a Newton–Raphson refinement, a combination that provides stability to the root finding. This analysis demonstrates that the distribution of the network edge lengths is directly reflected in the SIPR; in particular, we obtain that a small set of dominant peaks in the spectrum is more likely to appear when short fiber segments are comparatively abundant with respect to the average length. We also develop a more sophisticated categorization of the photonic networks based on machine learning (ML), further highlighting the correlation between the statistical properties of the network edge lengths and the ultimate emission features.
These results open an avenue for engineering the optical properties of photonic networks through controlled design, with significant potential for next
network-based photonic devices and intelligent systems.

\begin{figure}[ht!]
    \centering
    {\raggedright (a)\par}
    \includegraphics[width=0.65\linewidth]{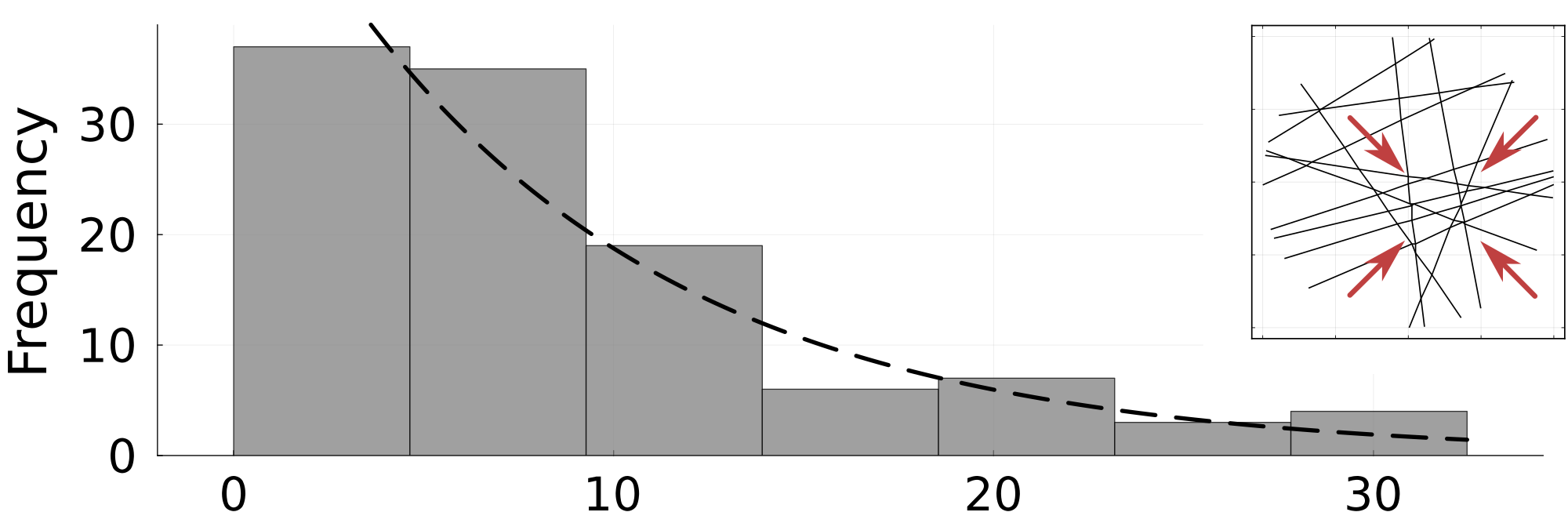}

    {\raggedright (b)\par}
    \includegraphics[width=0.65\linewidth]{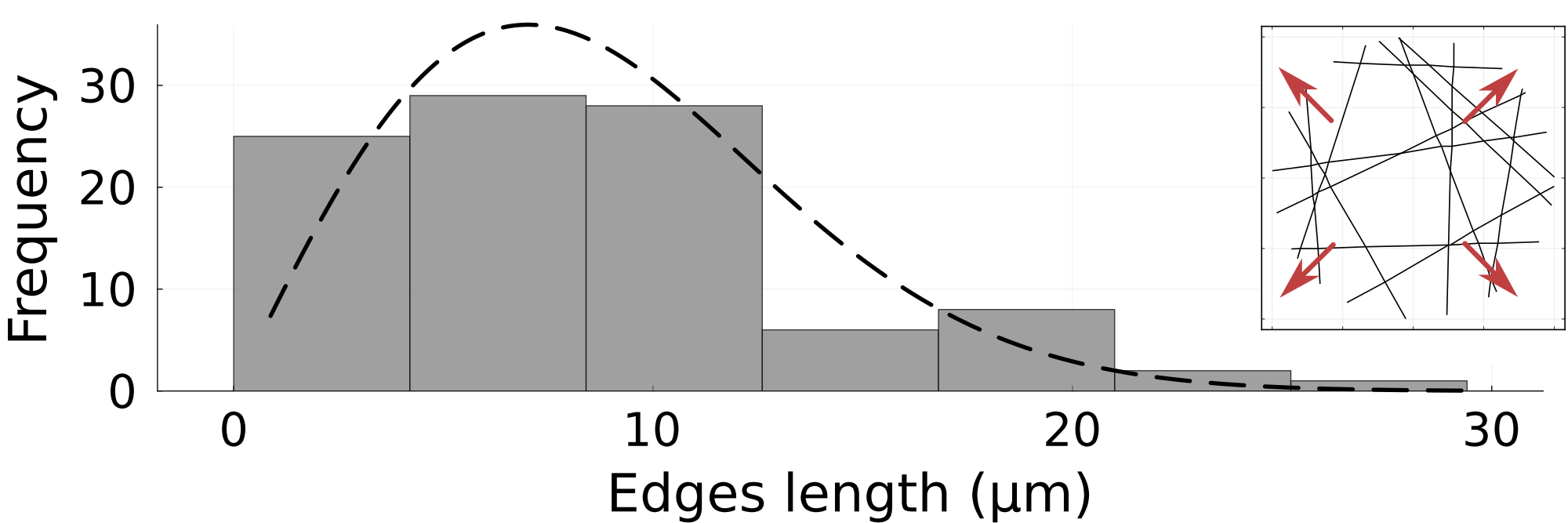}
    
    \caption{(a),(b) Distributions of the edge lengths for two different network configurations (insets), with Poisson (a) and Wigner-Dyson (b) edge lengths distribution (fits with dashed lines), respectively. Inset: the arrows highlight the suppression of short segments in the Wigner–Dyson network, in contrast to their abundance in the Poisson case.}
    \label{fig:ESdistribuzioni}
\end{figure}

\textit{Length Distribution of Network Edges.}-- First, we consider networks composed of fibers arranged randomly and focus on the \emph{crowding} of the fibers. The insets of Fig.~\ref{fig:ESdistribuzioni} show two examples of different crowding scenarios: (a) where higher crowding leads to a larger number of shorter fiber segments, and (b) where fibers avoid crowding and longer segments are present. Hence, we divide the networks into two sets: one in which the segments lengths followed a  Poisson distribution [like in (a)], and another where they follow a Wigner--Dyson distribution [like in (b)]. The Poisson distribution corresponds to uncorrelated fiber lengths, where short and long segments occur independently, while the Wigner--Dyson-like distribution exhibits fiber crossing  repulsion, suppressing the density of very short lengths. We point out that,  fiber crowding can be experimentally controlled through the process parameters and the electrospinning collector design \cite{Li2024}, while different and highly controlled geometries and distributions of fibers in emissive networks can be obtained experimentally by printing and near-field electrospinning methods~\cite{di2013near}.  To build the two sets of network configurations, we considered a set of 13 straight fibers randomly distributed, which results in multiple intersections.  A network area of radius $R = 40\,\mu\mathrm{m}$ is uniformly excited.  To account for the finite transverse size of the fibers---which allows for the formation of nodes connected to multiple edges---we discretize the space with a resolution of $0.5\,\mu\mathrm{m}$.
We then generate several fiber network samples, also classified into two categories: those whose fiber length distributions are closer to a Poisson distribution, and those resembling a Wigner--Dyson one (Fig.~\ref{fig:ESdistribuzioni}). The classification is performed using the Kolmogorov--Smirnov distance, which quantifies the maximum difference between the empirical cumulative distribution function (CDF) of the sample and the theoretical CDF~\cite{massey1951kolmogorov}, (see the Supplemental Material for more details). This approach resulted in two sets of configurations, with approximately 560 and 160 individual network configurations for Poisson and for Wigner-Dyson configurations, respectively, whose spectral properties were computed and statistically analyzed. 

\begin{figure}[ht!]
	\centering
    \includegraphics[width=0.8\linewidth]{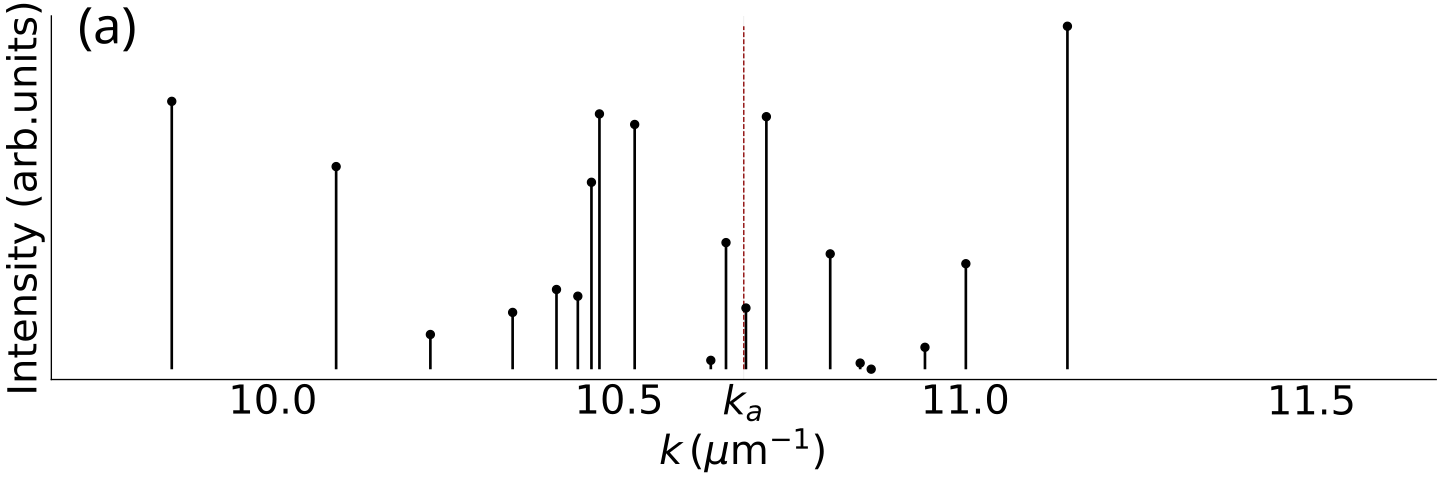}
	
    \vspace{4mm}
    \includegraphics[width=0.6\linewidth]{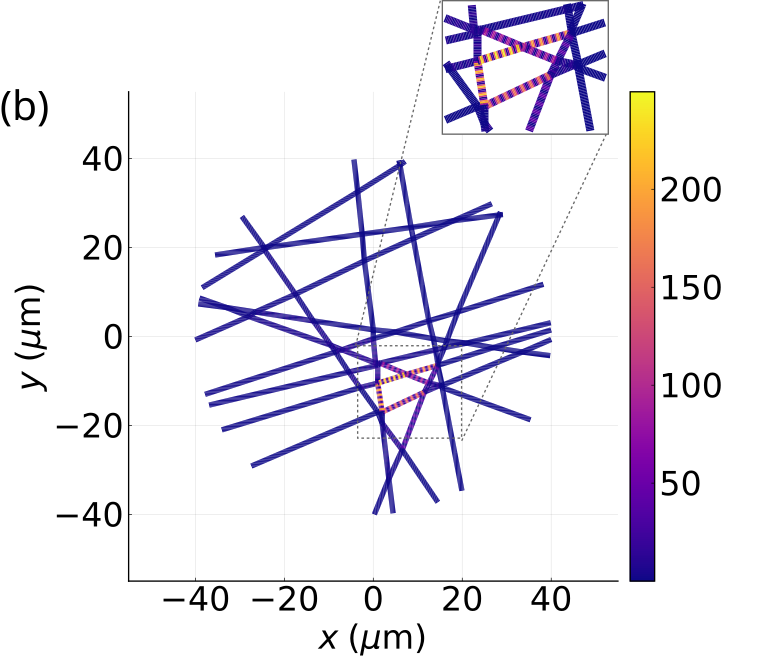}
    
	\caption{(a) Emission spectrum for a pump strength $D_0$ slightly above the gain-clamping threshold --- set here to 105\% of $0.044\times D_c$, which corresponds to the gain-clamping pump strength of the sample. The dashed line marks the position of the gain center.
   (b) Network geometrical configuration. The colorscale shows the spatial profile of the most intense mode at $D_0$, which is significantly localized. The colorscale shows the intensity of the electric field in $E_c^2$ units.}
	\label{fig:spect_modes}
\end{figure}

\textit{Theoretical Modeling.}---To calculate the NL spectrum, we consider the SALT equations, describing the steady-state behavior of lasers by coupling Maxwell’s equations with a gain medium modeled as a multi-level system, under the assumptions of time-harmonic fields and a stationary population inversion. SALT naturally accounts for the laser threshold -- the pump strength at which the gain balances the losses and a mode begins to lase. SALT also captures the so-called \emph{gain-clamping} transition: for pump values above a certain level, strong modal interactions suppress all other modes, preventing them from turning on, even as the pump strength increases~\cite{ge2010steady}. 
By assuming that the electric field is a superposition of modes $E(\mathbf{r}, t) = \sum_{\mu} \Psi_{\mu}(x) e^{-i \omega_{\mu} t}$, the SALT equation can be written as
\begin{multline}
	\label{eq:SALTI}
	\Bigl[ \nabla^2 + \bigl( n^2 + \frac{\gamma_\perp}{(\omega^\mu - \omega_a)+i\, \gamma_\perp}\\
	\times  \frac{D_0(x)/D_c}{1+\sum_\nu \Gamma_\nu | \Psi_\nu(x)|^2} \bigr) {k^\mu}^2
	\Bigr] \Psi_{\mu} = 0,
\end{multline}
where $D_c\equiv \epsilon_0\, \hbar\, \gamma_\perp/g^2$ is the natural reference scale for pumping strength (in the following we measure $D_0$ in units of $D_c$, i.e.~$D_c=1$ in the SALT equation), $k^\mu \equiv \omega^\mu/c$ and $\Gamma_\nu\equiv \Gamma(\omega^\nu)\equiv \gamma_\perp^2/[(\omega^\nu - \omega_{a})^2 + \gamma_\perp^2]$. Here $n=n(x,\omega^\mu)$ is the refractive index and the electric field is measured in units of $E_c\equiv \hbar\, \sqrt{\gamma_\parallel\, \gamma_\perp}/(2\, g)$. The wavenumber of the gain center (i.e., the characteristic wavenumber of the gain medium) is denoted by $k_a$, while $\gamma_{\parallel,\perp}$ are phenomenological damping constants, and $g$ is the dipole moment matrix element. Within each single-mode fiber, the SALT Eq.~\eqref{eq:SALTI} reduces to a one-dimensional ordinary differential equation in the spatial coordinate (see the Supplemental Material for more details). As parameters of the equation, we choose \( k_a = 10.68\ \mu\mathrm{m}^{-1} \), \( \gamma_\perp = 0.5\ \mu\mathrm{m}^{-1} \), (note that the equations depend on the longitudinal damping parameter $\gamma_{\parallel}$ only through the value of the natural electric-field unit $E_c$, and therefore the results obtained are independent of $\gamma_{\parallel}$
) and a refractive index \( n = 1.5 \) (non-dispersive regime), consistent with the experimental conditions of a network of fibers of polymethyl methacrylate doped with Rhodamine-6G dye ~\cite{saxena2022sensitivity}. Spectra are calculated in the interval $k\in [8,13]$ $\mu\text{m}^{-1}$.
A spectrum for a particular NL realization, for a pump strength value just above the gain clamping transition, is shown in Fig.~\ref{fig:spect_modes}, along with the spatial profile of the most intense mode, which appears to be fairly localized (cf. nonlocalizing random lasers~\cite{mujumdar2004amplified}).  In general, the lasing modes involve more than a single edge of the network.

\textit{Spectral Properties of the Networks}.---We use the numerical model to investigate the correlations between the network geometry and the spectral properties of the NL. To this aim, we numerically solve the nonlinear system of equations arising from SALT.

\begin{figure}[ht!]
	\centering
	{\raggedright (a)\par}
    \includegraphics[width=0.7\linewidth]{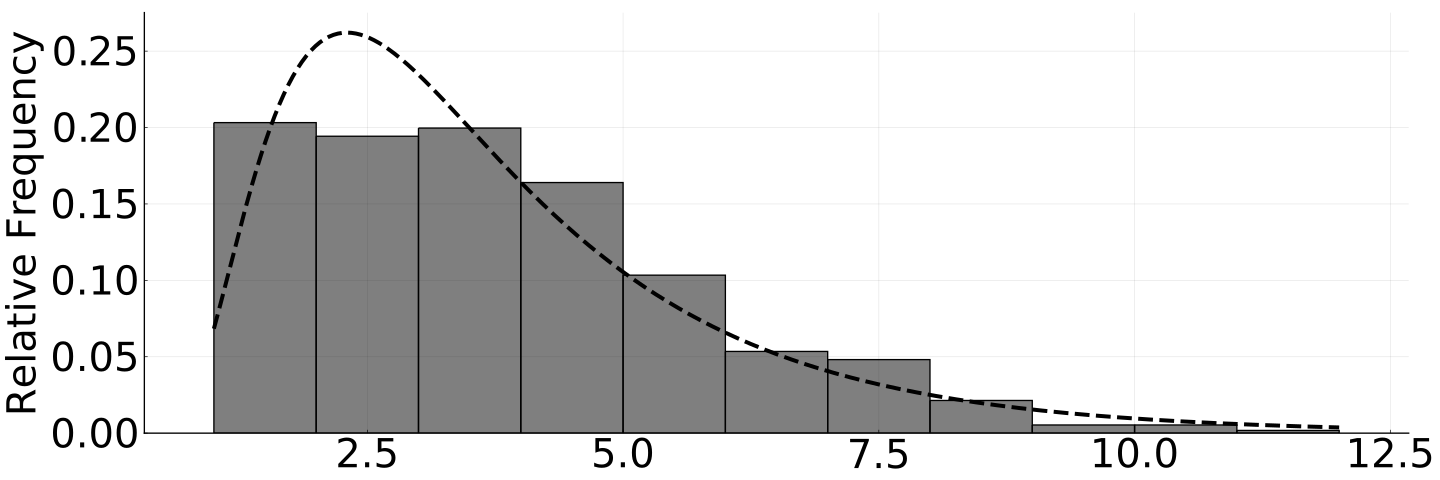}
    
	{\raggedright (b)\par}
    \includegraphics[width=0.7\linewidth]{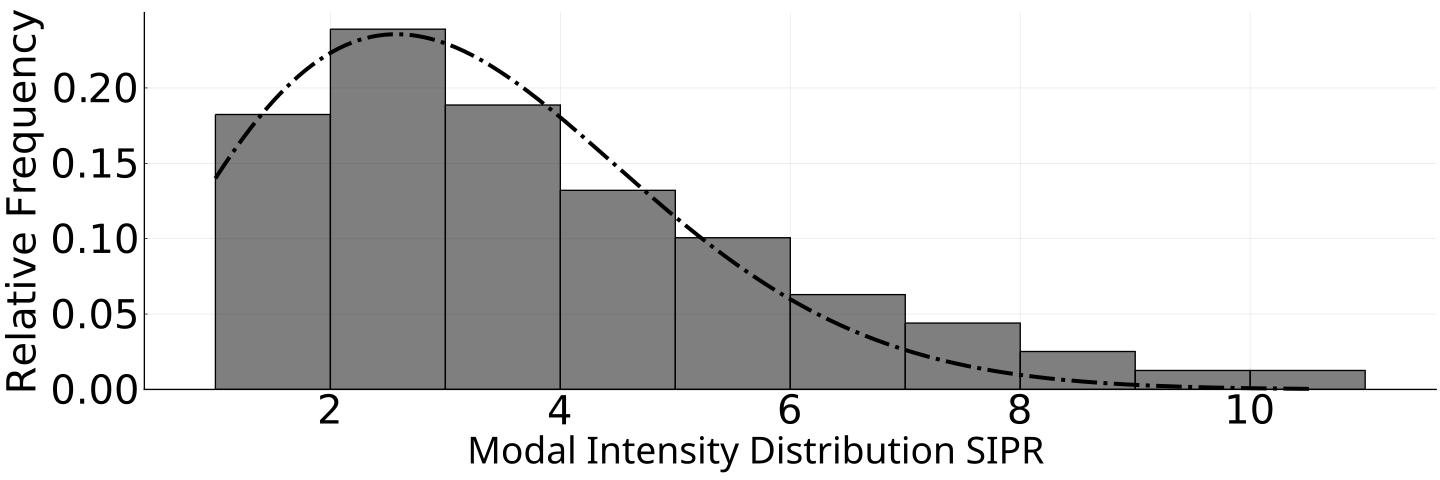}
	\caption{Distribution of the spectral inverse participation ratio (SIPR) of the modal intensity of the emission spectrum, for Poisson (a) and Wigner–Dyson (b) edge lengths distribution, respectively.  The dashed line in (a) and the dashed-dotted line in (b) are fit to the data by a Gamma--LogNormal and  Inverse Gaussian function, respectively. Examples of the length distributions of the fiber segments used in the calculations are shown in Fig.~\ref{fig:ESdistribuzioni}(a) and Fig.~\ref{fig:ESdistribuzioni}(b), corresponding to the Poisson and the Wigner–Dyson cases, respectively.}
	\label{fig:IPR}
\end{figure}

For a statistical analysis of the spectral properties, we introduce the spectral inverse participation ratio (SIPR), defined as
\begin{gather}
	    \text{SIPR} =
	\biggl[
	\sum_\mu  \left(  \frac{I_\mu}{\sum_\nu I_\nu} \right)^2
	\biggr]^{-1},
\end{gather}
where $I_\mu$ are the intensities of the calculated spectral lines. The SIPR equals 1 when the spectrum consists of a single mode, and approaches $N$ when the intensity is evenly distributed among $N$ modes. Analogously to the inverse participation ratio used for determining the spatial localization of modes~\cite{saxena2022sensitivity}, the indicator here introduced provides a measure of the spread of the emission in the wavelength space
and describes the distribution of intensity among modes (although different spectral configurations may yield similar values).
%, while detuning probes the spectral position relative to the gain center.}
SIPR is calculated for the set of networks analysed by the SALT theory, featuring either Poisson or Wigner--Dyson distributions of the network edge lengths. The results are shown in Fig.~\ref{fig:IPR}. The correlation between SIPR and fiber crowding is patent:  in the Poisson case, where short and long segments occur independently, the SIPR distribution is also Poisson-like, signaling the absence of an underlying spectral structure; surprisingly, in the Wigner-Dyson case, where very short segments are suppressed, low SIPRs are missing from the statistics, meaning that modal intensity distribution is more even.  This separation is quantitatively confirmed by $\chi^2$ tests against Inverse Gaussian and Gamma--LogNormal functions, clearly distinguishing the two cases in Fig.~\ref{fig:IPR} (see SM).

In addition we analyze the following spectral features, considering a pump strength just above the gain clamping transition: the intensity-weighted detuning with respect to the gain center wavenumber $k_a$ (IWD) and the detuning of the maximum intensity peak (MID), 
\begin{gather}
    \text{IWD} = \bigl(\sum_\mu k_\mu  \frac{I_\mu}{\sum_\nu I_\nu}\bigr) - k_a,\qquad
    \text{MID} = k_{\bar{\mu}} - k_a,
\end{gather}
where $\bar{\mu}$ is the wavevenumber corresponding to the maximum intensity peak.

\begin{figure}[ht!]
    \centering

    \begin{minipage}{0.45\linewidth}
        {\raggedright (a)\par}
        \includegraphics[width=\linewidth]{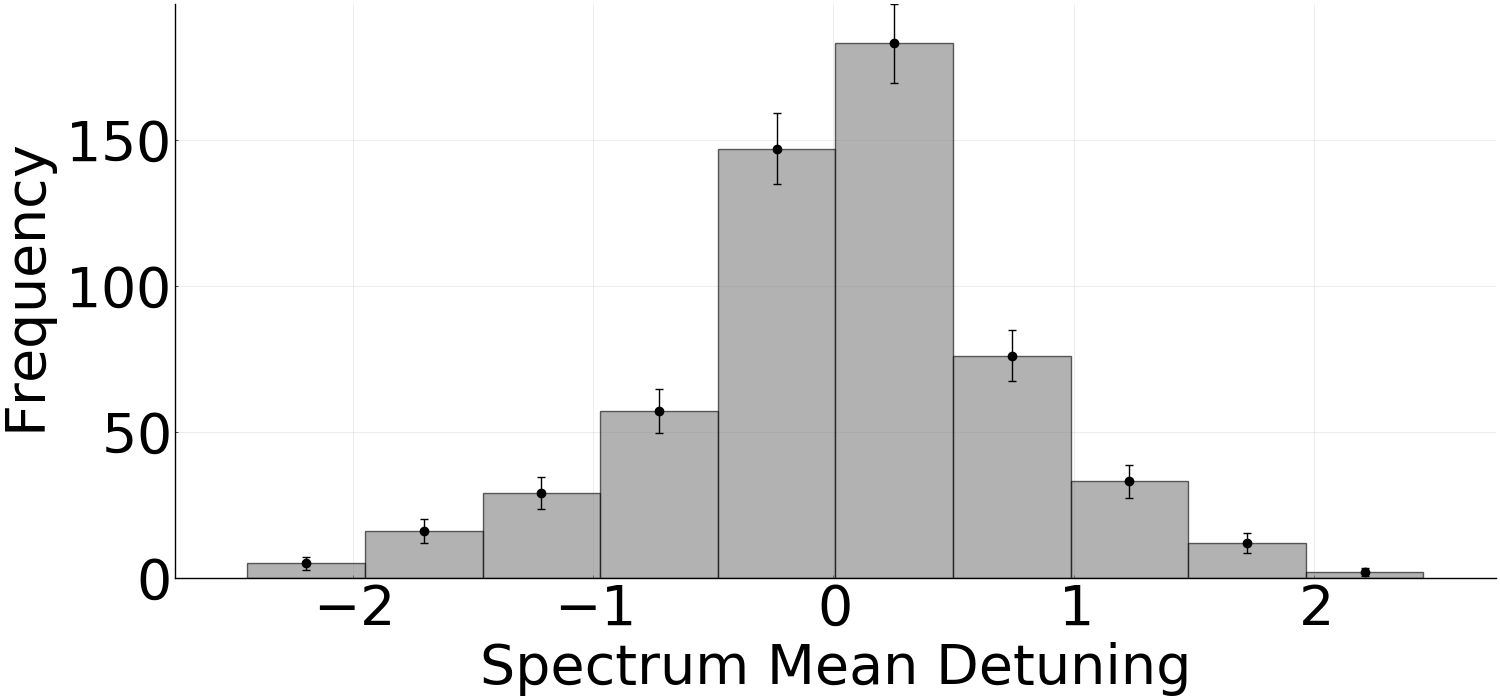}
    \end{minipage}
    \hfill
    \begin{minipage}{0.45\linewidth}
        {\raggedright (b)\par}
        \includegraphics[width=\linewidth]{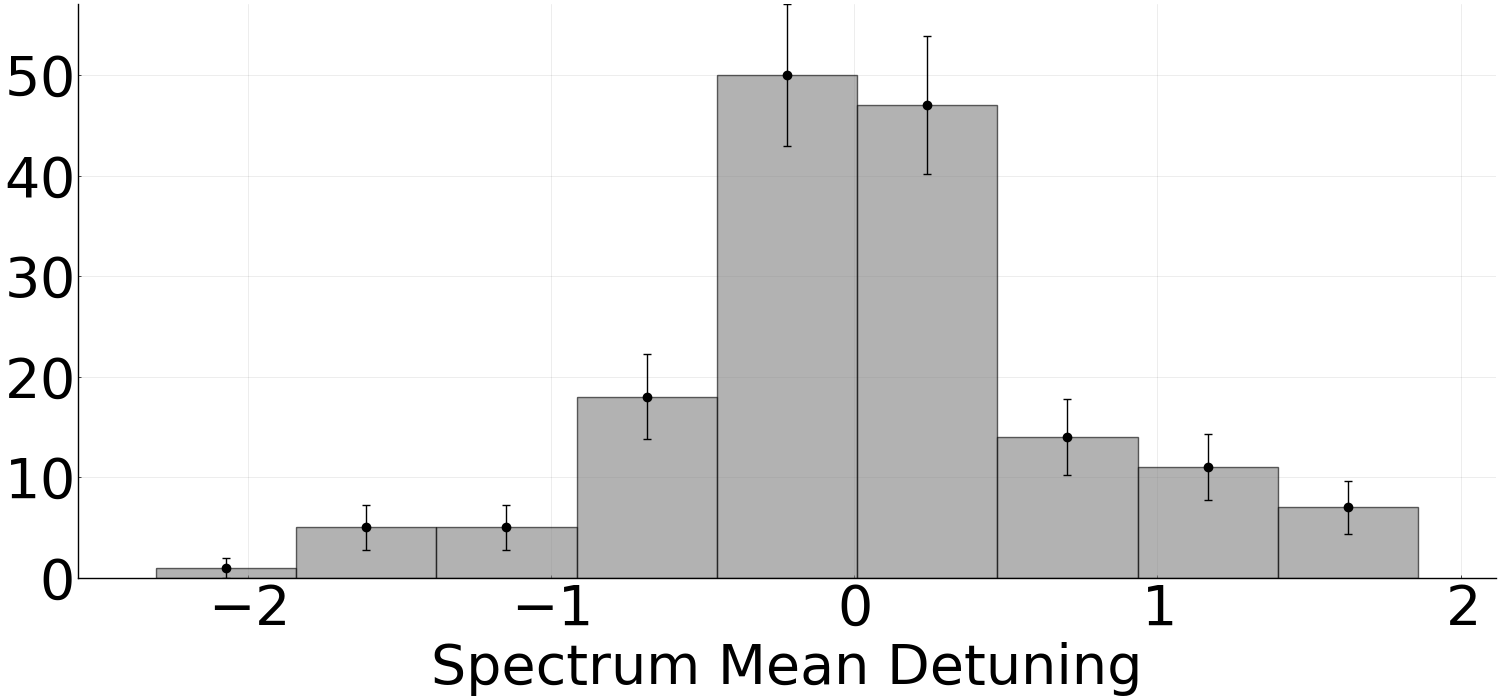}
    \end{minipage}

    \vspace{2ex}

    \begin{minipage}{0.45\linewidth}
        {\raggedright (c)\par}
        \includegraphics[width=\linewidth]{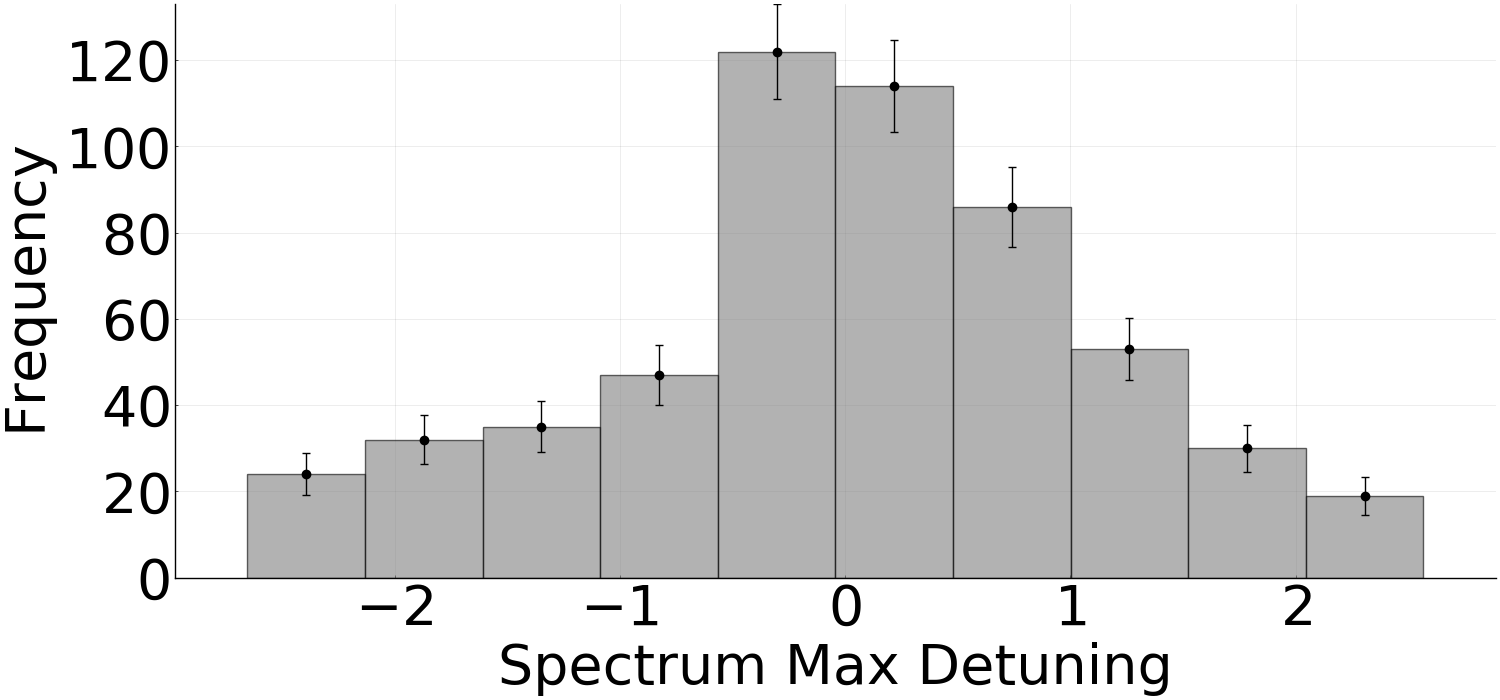}
    \end{minipage}
    \hfill
    \begin{minipage}{0.45\linewidth}
        {\raggedright (d)\par}
        \includegraphics[width=\linewidth]{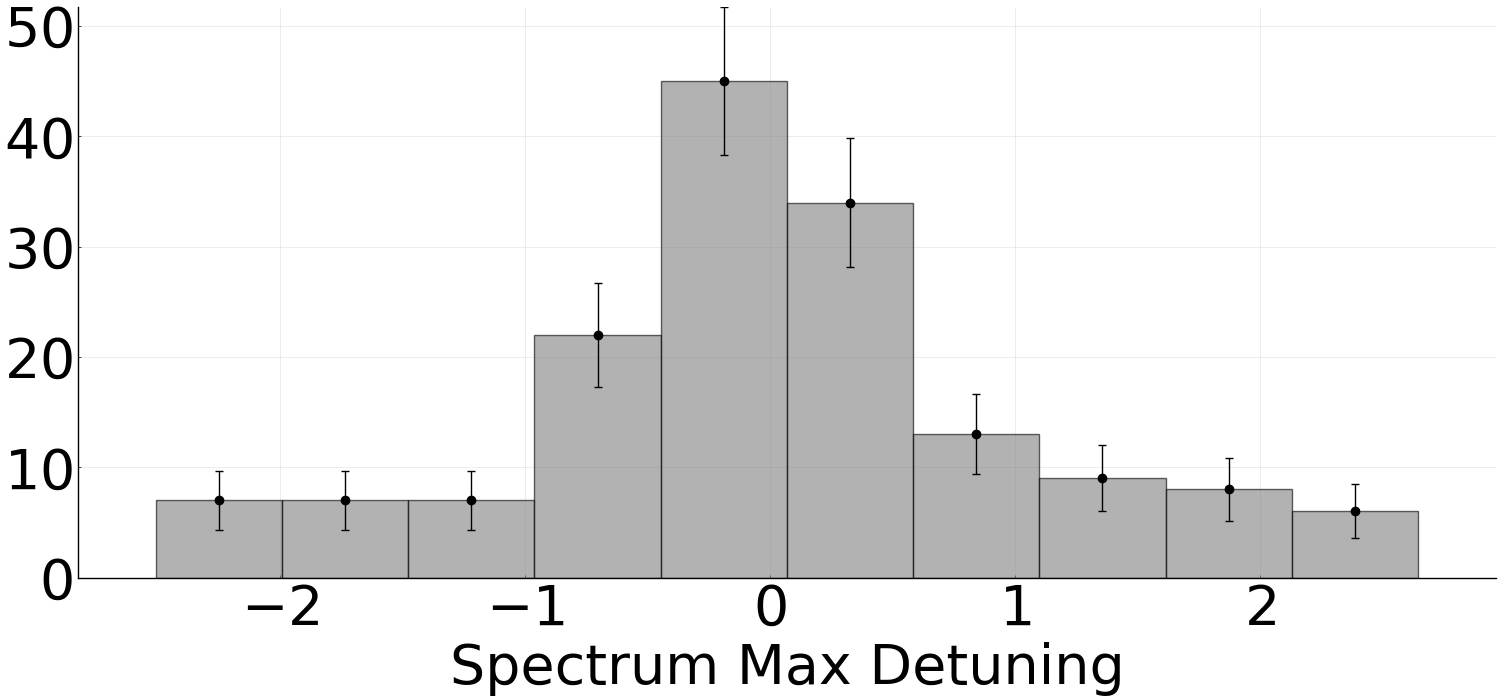}
    \end{minipage}

    \caption{(a),(b) IWD distributions for the Poisson (a) and Wigner--Dyson cases (b), respectively,  with error bars. (c),(d) MID distributions for the Poisson (c) and Wigner--Dyson cases (d), respectively. }
    \label{fig:detuning}
\end{figure}

Figure~\ref{fig:detuning} reports the obtained data along with the corresponding error bars, which highlight similar  behavior.

\begin{figure*} 
	\centering
	
	\subfloat[]{\includegraphics[width=0.24\linewidth]{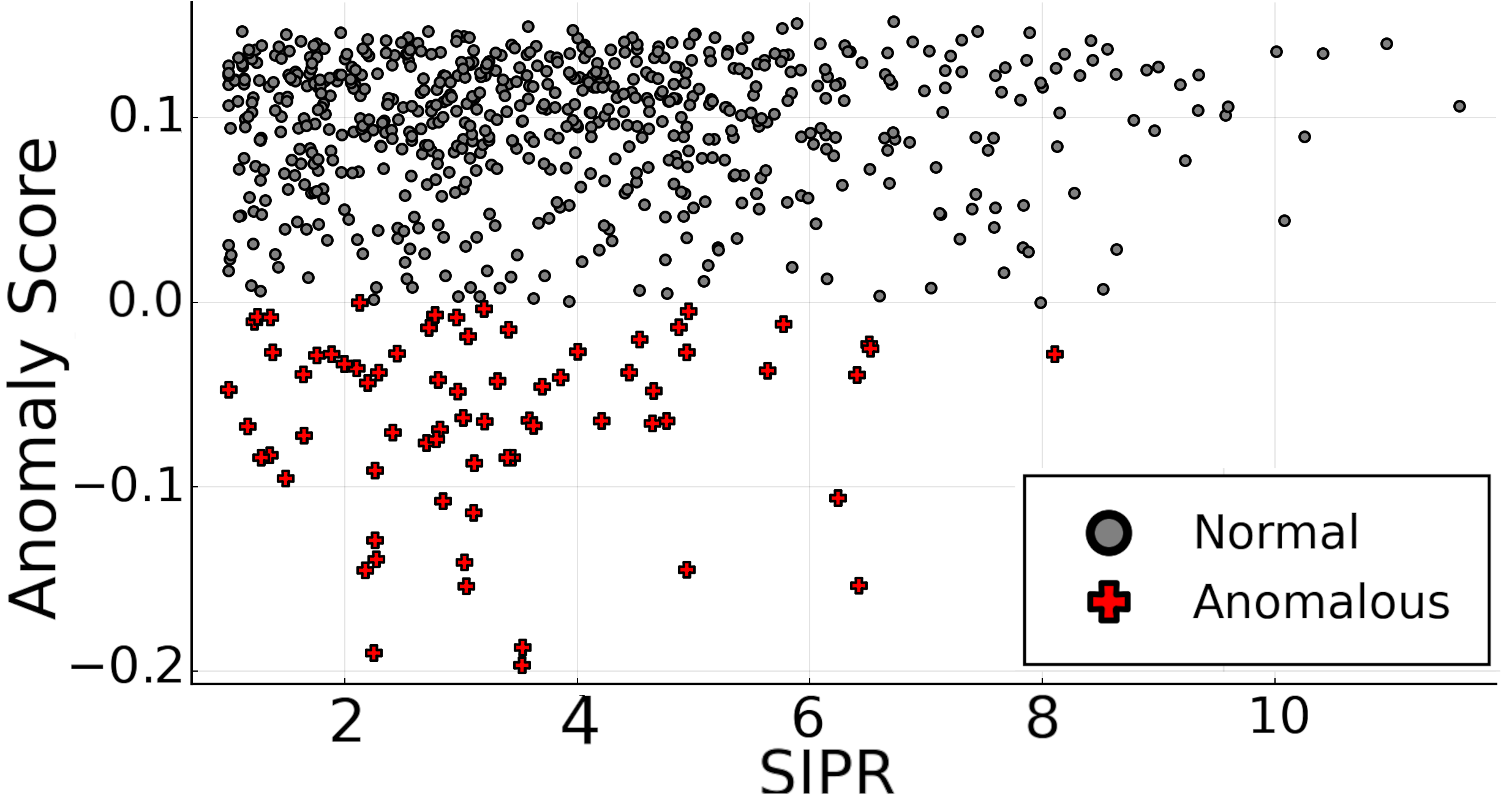}}
	\hfill
	\subfloat[]{\includegraphics[width=0.32\linewidth]{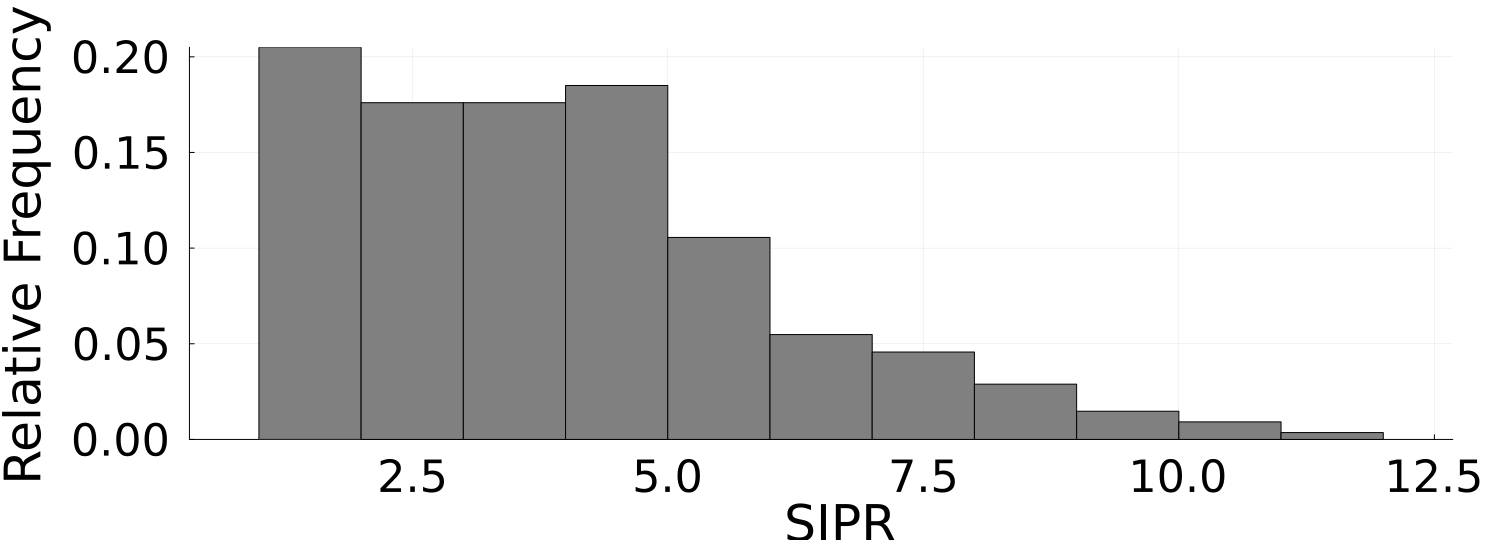}}
	\hfill
	\subfloat[]{\includegraphics[width=0.32\linewidth]{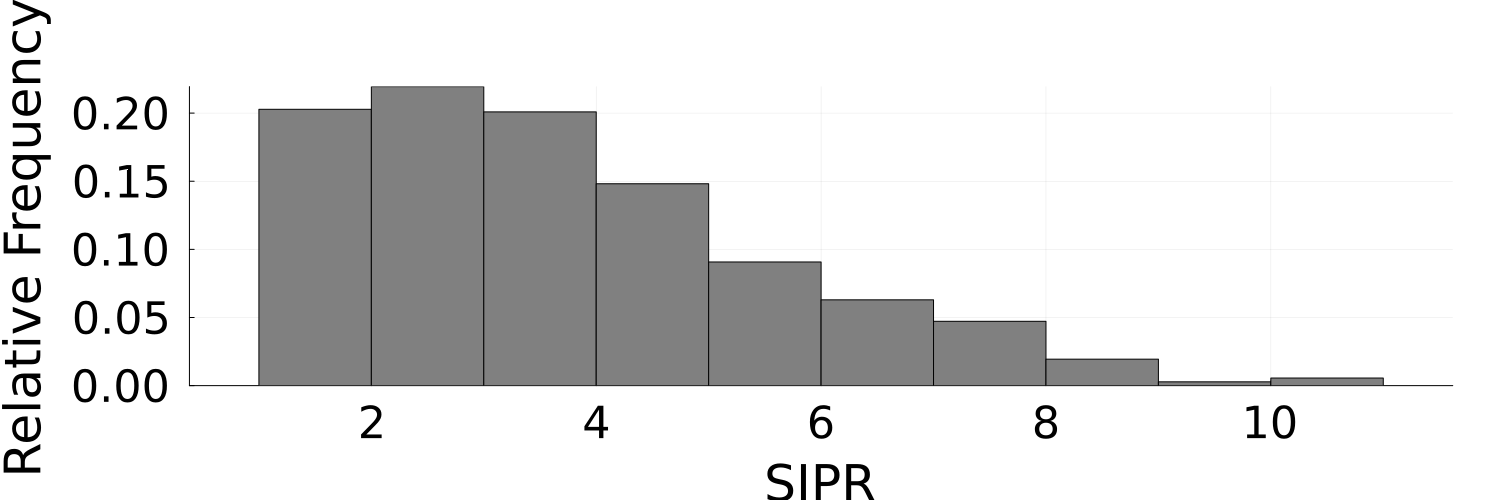}}
	\caption{Anomaly score of network realizations computed using the IFA algorithm as a function of the SIPR values. Circles (crosses) indicate normal (anomalous) network realizations, as defined in the main text. Panels (b) and (c) show the SIPR distributions for the normal and anomalous cases, respectively.}
	\label{fig:IFA}
\end{figure*}

\textit{Trails Network Analysis.}---Recently, neural networks have been successfully employed to visualize and control complex laser systems involving nonlinear mode interactions~\cite{ng2024mode}. Here, for each realization of the fiber network, we consider the possible trails. In graph theory, a trail is a sequence of adjacent edges in which no edge is repeated. In general, given the nonlinear nature of the SALT equations and the randomness of the network geometry in NLs, it is not straightforward to quantitatively estimate the influence of the number of nodes on the relevance of a trail to the spectral features. Hence, to gain a better understanding of the relationship between the network geometry and the emission spectrum properties, we analyze our system using ML techniques. We group trails according to their number of nodes, since we expect that the larger the number of nodes, the greater the optical losses associated with the trail and, consequently, the smaller its contribution to the resulting overall spectrum. This is further supported by the spatial localization of the emission of individual modes, as shown in Fig.~\ref{fig:spect_modes}(b).

In the following, we only consider trails with at most six nodes. We use the Isolation Forest algorithm (IFA)~\cite{liu2012isolation}, an unsupervised ML technique developed for anomaly detection, which operates by recursively partitioning the data using randomly constructed \emph{decision trees}. Moreover, unlike clustering methods such as K-Means—which perform well primarily on data distributed as Gaussian mixtures—IFA does not rely on any specific assumption about the underlying data distribution, making it particularly robust in highly heterogeneous settings like ours.

For each network realization (NR), we build a feature vector by concatenating the mean and standard deviation of trail lengths for each node count. This results in 12-dimensional feature vectors (6 means and 6 standard deviations), which are used as input to the model. For each NR, we compute the so-called \emph{anomaly score} [see Fig.~\ref{fig:IFA}(a)]. Lower, negative scores correspond to decision trees with more splits. We dub \emph{anomalous} the data points that fall below the 10th percentile.   The type of edge distribution of each NR affects its anomaly score, with the mean score for the Poisson distribution being approximately 26\% higher than that of the Wigner-Dyson distribution. The SIPR distributions for the normal and anomalous NR are shown in Fig.~\ref{fig:IFA}(b,c), respectively.
The anomalous NRs exhibit a suppressed occurrence of low SIPR values, as shown in Fig.~\ref{fig:IFA}(a,b).
%Furthermore, the mean SIPR is approximately 15\% higher for the outliers, while it remains comparable between Poisson and Wigner-Dyson fiber length distributions [Fig.~\ref{fig:IFA}(a)]. 
The mean SIPR of the normal NRs [Fig.~\ref{fig:IFA}(b)] is  roughly 15\% larger than that of the anomalous ones [Fig.~\ref{fig:IFA}(c)], with $\langle \mathrm{SIPR} \rangle_{\mathrm{norm}} = 3.86 \pm 0.09$
and $\langle \mathrm{SIPR} \rangle_{\mathrm{anom}} = 3.30 \pm 0.21$. This is  in stark contrast to the nearly identical values observed for the Poisson/Wigner–Dyson groups of the NRs  (not shown in the figure).
Overall, these results highlight a clear correlation between the spectral features and the distribution of trails.

In summary, the emission properties of networks of interconnected fibers is calculated by solving the SALT equations for a statistically significant number of networks, each with different geometric configurations and with either Poisson or Wigner-Dyson distribution of their edge lengths. By analyzing the spectral properties using the SIPR, a clear correlation was found with the specific distribution.
Cross-saturation and gain-clamping effects associated with spatial overlap of modes and consequently with the specific network topology underneath can be mechanisms that closely correlate the distributions featured by the physical network and those describing the ultimate optical response.
In addition, we showed that unsupervised anomaly detection via IFA can effectively distinguish between spectral regimes in network lasers based on geometrical characteristics. 
In the SM, we further extend our analysis by investigating the statistical properties of fiber segment lengths using a continuous descriptor, namely the Brody distribution.
These results provide a novel statistical framework for characterizing and ultimately controlling the lasing behavior in NLs.

\textit{Acknowledgment} We gratefully acknowledge the National Centre for HPC, Big Data and Quantum Computing, under the National Recovery and Resilience Plan (NRRP), Mission 4 Component 2 Investment 1.4 CUP I53C22000690001 funded from the European Union-NextGenerationEU.

\bibliography{NLs.bib}

\end{document}

\title{Supplemental material for "Statistical Insight into the Correlation of Geometry and Spectral Emission in Network Lasers"}
	
	\author{Camillo Tassi}
	\email{camillo.tassi@gmail.com}
	\affiliation{Dipartimento di Fisica, Università di Pisa, Largo Bruno Pontecorvo 3, I-56127 Pisa, Italy}
	
	\author{Riccardo Mannella}
	\affiliation{Dipartimento di Fisica, Università di Pisa, Largo Bruno Pontecorvo 3, I-56127 Pisa, Italy}
	
	\author{Andrea Tomadin}
	\affiliation{Dipartimento di Fisica, Università di Pisa, Largo Bruno Pontecorvo 3, I-56127 Pisa, Italy}
	
	\author{Andrea Camposeo}
	\affiliation{NEST, Istituto Nanoscienze - CNR and Scuola Normale Superiore, Piazza San Silvestro 12, I-56127 Pisa, Italy}
	
	\author{Dario Pisignano}
	\affiliation{Dipartimento di Fisica, Università di Pisa, Largo Bruno Pontecorvo 3, I-56127 Pisa, Italy}
	\affiliation{NEST, Istituto Nanoscienze - CNR and Scuola Normale Superiore, Piazza San Silvestro 12, I-56127 Pisa, Italy}

	%\date{}

\begin{abstract}
In this Supplemental Material File we describe the method used to classify the network laser (NL) realizations, based on the network edge length distribution; we describe the numerical method used to solve the Steady--State ab Initio Laser Theory (SALT) equation for NL system in the Single Pole Approximation (SPA) and we show how we test it; we complement the statistical analysis provided in the main text. 
\end{abstract}

\maketitle

\section{Classification of the Realizations Based on Fiber Lengths}

In the case of fibers modeled as lines filling a two-dimensional plane, the distribution of segment lengths between two nodes follows a Poisson distribution. We also consider network realizations that exhibit edge-length distributions that are phenomenologically closer to Wigner–Dyson statistics, since such configurations can in principle be realized. Indeed fiber geometry can be experimentally controlled through the process parameters and the electrospinning collector design, and different and highly controlled geometries and distributions of fibers in emissive networks can be obtained experimentally~\cite{di2013near}.

We analyzed several realizations of straight fibers randomly placed within an excited area of radius $R = 40\,\mu\mathrm{m}$ and with multiple intersections. To account for the finite transverse size of the fibers—which allows the formation of nodes with multiple connections—we discretized the space with a resolution of $0.5\,\mu\mathrm{m}$.

The realizations were classified into two distinct categories: those whose fiber length distributions are closer to a Poisson distribution, and those more consistent with a Wigner--Dyson distribution.

The classification was based on the Kolmogorov--Smirnov distance, which quantifies the maximum deviation between the empirical cumulative distribution function of the data and the corresponding theoretical cumulative distribution function~\cite{massey1951kolmogorov}. In particular, given a set of fiber lengths $\{\ell_i\}_{i=1}^n$, we first normalized them by the mean:
\begin{gather}
x_i = \frac{\ell_i}{\langle \ell \rangle}, \quad \text{with } \langle \ell \rangle = \frac{1}{n} \sum_{i=1}^n \ell_i.
\end{gather}
The normalized lengths were sorted in increasing order, and the empirical cumulative distribution function was defined as:
\begin{gather}
F_{\mathrm{emp}}(x_i) = \frac{i}{n}, \quad i = 1, \dots, n.
\end{gather}

We compared $F_{\mathrm{emp}}$ to two theoretical models:
\begin{itemize}
  \item the exponential cumulative distribution function associated with a Poisson (uncorrelated) process:
  \begin{gather}
  F_{\mathrm{Poisson}}(x) = 1 - e^{-x},
  \end{gather}
  \item and the Wigner--Dyson distribution, characteristic of level repulsion:
  \begin{gather}
  F_{\mathrm{WD}}(x) = 1 - \exp\left(-\frac{\pi x^2}{4}\right).
  \end{gather}
\end{itemize}

The corresponding Kolmogorov--Smirnov distances were then computed as:
\begin{gather}
D_{\mathrm{Poisson}} = \max_i \left| F_{\mathrm{emp}}(x_i) - F_{\mathrm{Poisson}}(x_i) \right|, \quad
D_{\mathrm{WD}} = \max_i \left| F_{\mathrm{emp}}(x_i) - F_{\mathrm{WD}}(x_i) \right|.
\end{gather}

Each sample was assigned to the distribution yielding the smaller distance. This allowed us to quantitatively identify which statistical model better describes the structure of each network realization.

 \begin{figure}[t]
     \centering
     \includegraphics[width=0.99\linewidth]{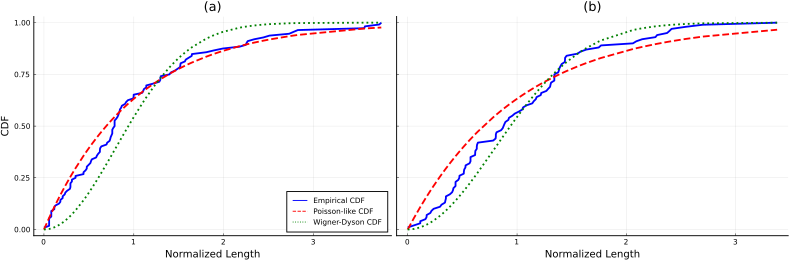}
     \caption{Panel~(a) shows the fiber length distribution  (blue line) corresponding to Fig.~\ref{fig:fibers}(a), which is closer to a Poissonian distribution. Panel~(b), on the other hand, associated with Fig.~\ref{fig:fibers}(b), displays a distribution (blue line) that more closely resembles the Wigner--Dyson case. The red and green dashed lines are the theoretical Poisson and Wigner--Dyson distributions respectively.}
     \label{fig:PVSWD}
 \end{figure}

 As an example of such analysis, the empirical cumulative distribution function was plotted alongside the two theoretical curves in Fig.~\ref{fig:PVSWD} corresponding to the two realizations in Fig.~\ref{fig:fibers}, respectively.

\section{Numerical Methods}

In order to solve the SALT equation as defined in the \textit{Theoretical Modeling} section of the main text:
\begin{gather}
	\label{eq:SALTI}
	\Biggl[ \nabla^2 + \biggl( n^2(x,\omega^\mu) + \frac{\gamma_\perp}{(\omega^\mu - \omega_a)+i\, \gamma_\perp}
	  \frac{D_0(x)/D_c}{1+\sum_\nu \Gamma_\nu \lvert \Psi_\nu(x)\rvert^2} \biggr) {k^\mu}^2
	\Biggr] \Psi_{\mu}(x) = 0,
\end{gather}
where $\Psi_\mu$ and $k^\mu\equiv \omega^\mu/c$ are unknown, for the single mode fibers of a network laser we considered the single pole approximation. 

This is a system of nonlinear equations, whose modal interaction term is given by:
\begin{gather}
    \label{eq:burning-hole}
    H_\text{nl}(x) \equiv \sum_\nu \Gamma_\nu \lvert \Psi_\nu(x)\rvert^2,
\end{gather}
also known in literature as the burning hole term~\cite{ge2010steady}.

To solve the equations, we considered the following three steps.

\subsection{Quasi-bound modes}

In the first step, we computed the quasi-bound modes: these are electromagnetic solutions of SALT Eq.~\eqref{eq:SALTI} in the absence of gain (i.e.~$D_0=0$), satisfying outgoing boundary conditions at infinity. They are non-Hermitian modes and the imaginary part of $k^\mu$ is negative, meaning that they grow exponentially at large distances~\cite{tureci2006self}.

Only modes sufficiently close to the gain center frequency \( k_a \) can contribute to the spectrum~\cite{ge2010steady}. Accordingly, we analyzed a region around \( k_a \) with width \( \Delta k \sim 12 \gamma_\perp \). To first approximation, the distance of the modes \( k^\mu \) from the real axis is proportional to the pump threshold required to activate the mode~\cite{gaio2019nanophotonic}; given the mode closest to the real axis, we included in our analysis all modes whose distance was significantly larger.

\begin{figure}[t]
    \centering
    \includegraphics[width=0.8\linewidth]{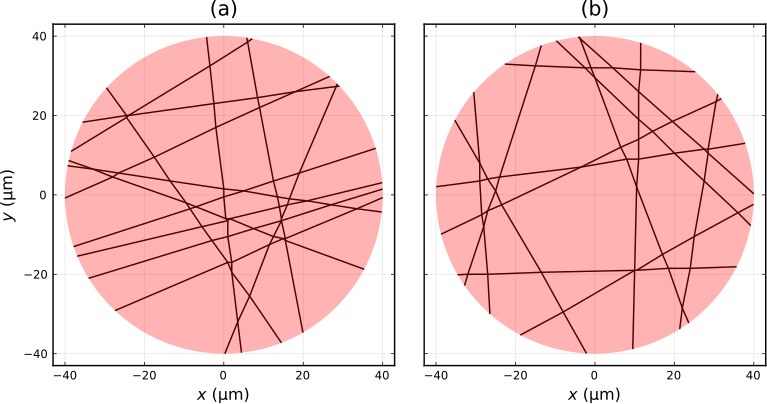}
    \caption{Network lasers composed of active single-mode waveguides. Interconnection points are nodes. The red spot of radius $R=40\, \mu$m indicates the excited region. Both samples consist of 13 fibers arranged in a random configuration.}
    \label{fig:fibers}
\end{figure}

Fig.~\ref{fig:fibers} shows two examples of network configurations obtained with randomly distributed fibers. Within each single-mode fiber, the SALT Eq.~\eqref{eq:SALTI} reduces to a one-dimensional ordinary differential equation in the spatial coordinate. As parameters of the equation, we have chosen \( k_a = 10.68\ \mu\mathrm{m}^{-1} \), \( \gamma_\perp = 0.5\ \mu\mathrm{m}^{-1} \), and a refractive index \( n = 1.5 \) (non-dispersive regime), consistent with the experimental conditions reported in Ref.~\cite{saxena2022sensitivity}, where networks of dye-doped polymer nanofibers were electrospun using a solution of polymethyl methacrylate and Rhodamine-6G (1\% dye-to-polymer weight ratio) in a mixture of chloroform and dimethyl sulfoxide.

In our case, for the fiber connecting node $i=1,2,\ldots$ to node $j$ (the nodes are fiber junctions), the quasi-bound modes take the form $\eta_{ij}(x) = \lambda_{ij}^+\, \mathrm{e}^{i k n x} +
\lambda_{ij}^-\, \mathrm{e}^{\mathrm{i} k n (l_{ij} - x)}$, where $l_{ij}$ is the fiber length; outgoing boundary condition was imposed on the fibers connected to nodes lying outside the pumped region: $\eta_{ij}(x) = \lambda_{ij}^+\, \text{e}^{i k n x}$, where we assume positive $x$ for $i<j$. At the fiber junctions, the electric field solutions are obtained by imposing the continuity of the field and the conservation of charge. The continuity condition reads $\eta_i \equiv \eta_{ij}(x=0) = \eta_{ki}(x=l_{ki})$, for all $j,\, k$ connected to node $i$, while the conservation of charge, following from the condition $\nabla \cdot \mathbf{E} = 0$, leads to 
$\sum_{j \colon j \leftrightarrow i} \left.\frac{d \eta_{ij}}{dx}\right|_{x=0}
= \sum_{k \colon k \leftrightarrow i} \left.\frac{d \eta_{ki}}{dx}\right|_{x=l_{ki}}$, 
where $j \colon j \leftrightarrow i$ are the nodes connected to node $i$ by fibers with $i < j$, and $k \colon k \leftrightarrow i$ are those with $k < i$. This gives rise to a system of equations of the form $L(k)\, \mathbf{x} = \mathbf{0}$, with 
\begin{gather}
\mathbf{x} \equiv
\begin{pmatrix}
\lambda_1^+ & \cdots 
& \lambda_{N_\text{int}+N_\text{ext}}^+ & \lambda_1^- & \cdots & \lambda_{N_\text{int}}^-
\end{pmatrix}^\top
\end{gather}
where a single index is assigned to each edge, $\lambda_{ij} \to \lambda_k$. Here, $N_\text{int}$ denotes the number of edges connecting two nodes within the pumped region, while $N_\text{ext}$ counts the edges connecting a node inside the spot to one outside. The matrix $L(k)$ is square, since for a graph the identity $\sum_{v \in V} d(v) = 2E$ holds, where $V$ is the set of vertices (or nodes), $d(v)$ is the degree of vertex $v$ (i.e., the number of edges incident to $v$), and $E$ is the total number of edges. The quasi-bound modes are therefore obtained by requiring that $\det L(k) = 0$.

It is worth noting that the approach used in Ref.~\cite{saxena2022sensitivity}, referred to as net-SALT, also relies on the single-pole approximation, as does the formulation adopted here. Compared to net-SALT, however, our matrix $L(k)$ has a larger dimension than the one defined there for computing quasi-bound modes, although it is also sparser. In our formulation, in particular, $\det L(k)$ has no poles. This allowed us to compute its zeros by first applying the Argument Principle (the number of zeros inside the contour $\Gamma$ is given by $1/(2\pi i) \oint_\Gamma f'(k)/f(k)\, dk)
$ to identify regions containing a single zero, followed by a Newton--Raphson refinement.
In Fig.~\ref{fig:modes}, the quasi-bound modes are shown for the network depicted in Fig.~\ref{fig:fibers}.

\subsection{Threshold Lasing Modes}

In the second step, we determined the threshold lasing modes by incorporating the gain medium while neglecting the modal interaction term~\eqref{eq:burning-hole}. This results in modes with real wavenumbers $k^\mu$.

Each threshold lasing mode $\phi_\mu(x)$ is obtained starting from a quasi-bound mode $\eta_\mu(x)$, by solving Eq.~\eqref{eq:SALTI} while neglecting the modal interaction term -- burning-hole term defined in Eq.~\eqref{eq:burning-hole} -- and gradually increasing $D_0$ (with $D_0 = 0$ corresponding to the quasi-bound mode) in small steps until condition $\Im k^\mu = 0$ is reached. As $D_0$ increases, the real part of $k^\mu$ slightly shifts towards the atomic transition frequency, as expected~\cite{ge2010steady}.

\begin{figure}[t]
    \centering
    \includegraphics[width=0.99\linewidth]{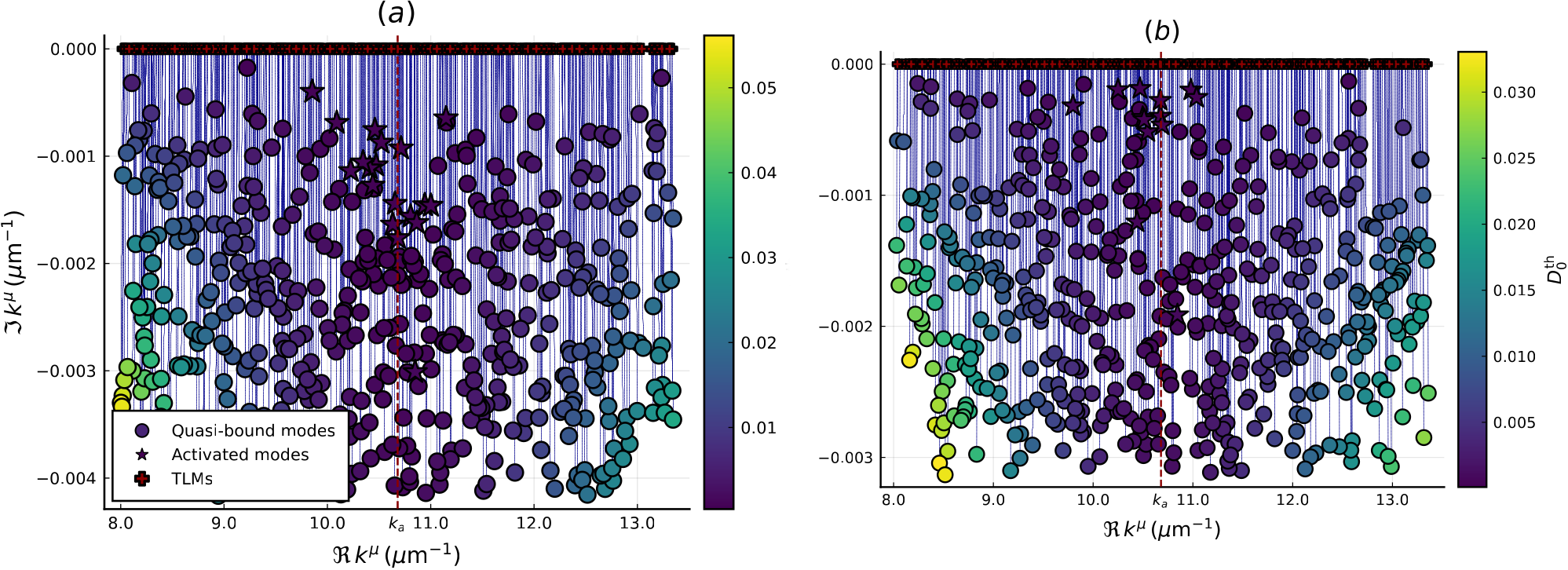}
    \caption{The quasi-bound modes have a negative imaginary part. However, only a subset of them actually contributes to the spectrum and can be activated, due to the gain-clamping transition. The threshold lasing modes are obtained from the quasi-bound modes by increasing $D_0$ (starting from $D_0 = 0$) and neglecting the burning-hole term, until the real axis is reached. The corresponding values $D_0^\text{th}$ represent the pump thresholds in the absence of mode competition $H_{\text{nl}}$, that is, assuming non-interacting modes. Panels (a) and (b) correspond to the two cases shown in Fig.~\ref{fig:fibers}, respectively.
}
    \label{fig:modes}
\end{figure}

In order to calculate the threshold lasing modes, we have to consider the substitution 
$n^2 \to n^2 + \gamma_\perp\, D_0/[(\omega^\mu - \omega_a) + \mathrm{i}\, \gamma_\perp ]$ 
in the SALT Eq.~\eqref{eq:SALTI}. In particular, as $D_0$ increases starting from zero in small steps, we use again the Newton-Raphson technique. Figure~\ref{fig:modes} illustrates the procedure that connects the quasibound modes to the threshold lasing modes (TLMs).

In Fig.~\ref{fig:modes}, the trajectories of the modes in the complex $k$-plane are depicted as the pump strength $D_0$ increases. These paths illustrate the transition from quasibound modes, characterized by complex wavenumbers with negative imaginary parts, to threshold lasing modes with real wavenumbers. This progression reflects the activation of lasing modes as the gain compensates for losses in the system.

\subsection{Single Pole Approximation}

In the final step, the lasing mode intensities were computed by solving the full SALT equation~\eqref{eq:SALTI} (including the burning hole term), under the assumption that the spatial profiles of the interacting modes remain fixed and equal to those of the corresponding TLMs: $\Psi_\mu(x) \simeq a_\mu\, \phi_\mu(x)$. The approximation is valid in high-Q cavities, because the scattering of the gain medium is much weaker than the scattering from the cavity itself~\cite{ge2010steady}.

 In order to obtain the amplitude $a^\mu(D_0)$ and the interacting thresholds $D^\mu_{0\text{int}}$ within this approximation, the SPA-SALT (single-pole approximation) equation reads~\cite{ge2010steady}:
 \begin{subequations}
 \begin{gather}
 	\frac{D_0}{D^\text{th}_{0\mu}} - 1 = \sum_\nu \Gamma_\nu\, \chi_{\mu\nu}\, I_\nu
 	\equiv \sum_\nu A_{\mu\nu}\, I_\nu,\\
 	\chi_{\mu\nu} \equiv \int_\mathcal{B} dx'\, n^2\, \phi_\mu^2(x')\, \lvert \phi_\nu(x')\rvert^2,\quad
 	I_\nu\equiv \lvert a^\mu\rvert^2,
 \end{gather}
 \end{subequations}
 where both the intensities $I_\nu$ and the interacting thresholds $D^\mu_{0,\text{int}}$ are unknown. In this equation, all the terms except $\chi_{\mu\nu}$ are real; this is consistent with the SPA, since for high $\mathcal{Q}$ the imaginary part of $\chi_{\mu\nu}$ must be small. We then put $\chi_{\mu\nu} \to \Re \chi_{\mu\nu}\geq 0$ in the SPA-SALT.
 
 The SPA-SALT is obtained by assuming that $a^\mu\neq 0$ and its solution require the following procedure~\cite{ge2010steady}: the first mode involved is the TLM associated with the smallest non-interacting threshold (indeed, as long as no mode is active, the hole-burning term remains zero). As the pump strength increases, additional modes can be activated.
 As part of a recursive argument, we assume that for a given value of $D_0$, $N$ modes are involved, labeled by $\mu = 1, \ldots, N$.
We can then solve the SPA-SALT equation, which shows that the lasing intensities increase linearly with the pump strength $D_0$.
 In order to find the interacting threshold for the next $N+1$ mode, we can observe that, for $D_0= D_{0,\text{int}}^{N+1}$ the SPA-SALT equation must give the same $I_1,\ldots,I_N$ values of  $D_0= D_{0,\text{int}}^{N+1}-0^+$, plus the solution $I_{N+1}=0$. In particular, if we consider $M$ possibly modes in total, the SPA-SALT gives
 \begin{subequations}
 \label{eq:SPA-SALT}
 \begin{gather}
 	D^\mu_{0, \text{int}} = \frac{D_0^\mu\left( 1-\sum_{\nu=1}^N A_{\mu\nu}\, b_\nu\right)}{1-D_0^\mu \sum_{\nu=1}^N A_{\mu\nu}\, c_\nu},\quad
 	\mu=N+1,\ldots, M,\\
     b_\mu \equiv \sum_{\nu=1}^N A^{-1}_{\mu\nu},\quad
     c_\mu \equiv \sum_{\nu=1}^N \frac{A^{-1}_{\mu\nu}}{D^\nu_0}
 \end{gather}
\end{subequations}
 the lowest $D_{0,\text{int}}^\mu$ gives the correct $N+1$ mode and the actual  $D_{0,\text{int}}^{N+1}$ threshold. The procedure can be iterated for all $\mu=1,\ldots,M$ modes considered.
 
As the pump strength increases, the matrix $A$ may eventually become singular; then, all higher thresholds are pushed off to infinity and no additional modes can turn on for any value of the pump, giving rise to the gain-clamping transition. It is also possible that, as the pump strength increases, some modes are turned off~\cite{ge2010steady}. Instead, in the gain-clamping regime considered in this work, no additional modes turn on or off upon increasing the pump strength; within the single-pole approximation, the intensities of the active modes grow linearly with the pump.

The calculation involves the inversion of a relatively small matrix -- Eq.~\eqref{eq:SPA-SALT} -- and the computation time is negligible once the TLMs have been determined.

\begin{figure}[t]
    \centering
    \includegraphics[width=0.99\linewidth]{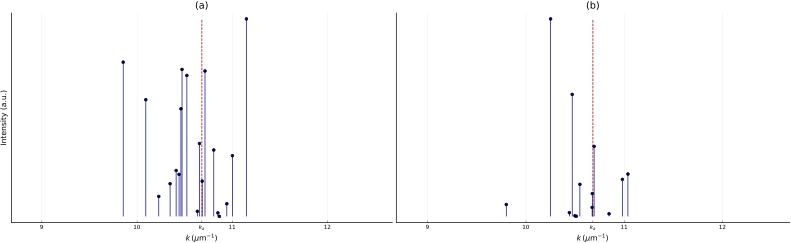}
    \caption{Spectrum for a pump strength $D_0$ just above the gain clamping value ($D_0$ at 105\% of the gain clamping value), corresponding to the two NL configurations shown in Fig.~\ref{fig:fibers}, respectively.}. 
    \label{fig:spect}
\end{figure}

In Fig.~\ref{fig:spect}, we show the spectrum for the two samples presented in Fig.~\ref{fig:fibers}, respectively. In this case, the detuning of the Wigner-Dyson-like realization is smaller compared to that of the Poisson realization. Indeed, it is important to emphasize that our analysis is based on the statistical properties of a large ensemble of configurations.

\subsection{Numerical Test}

\begin{figure}[t]
    \centering
    \includegraphics[width=0.87\linewidth]{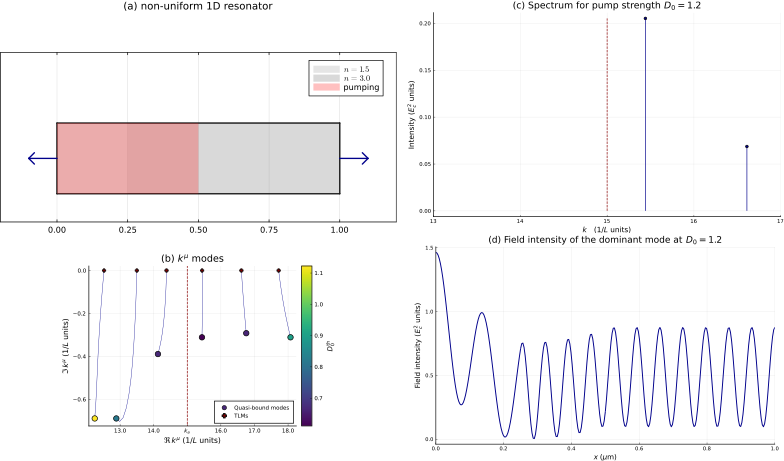}
    \caption{(a) Schematic representation of a one-dimensional nonuniform open resonator. The cavity extends from $x = 0$ to $x = L$ (lengths are measured in units of $L$), with open boundaries at both ends. Arrows indicate the outgoing boundary conditions at the edges. In panel (b), the quasi-bound modes and the threshold lasing modes are shown. Panel (c) displays the spectrum for a pump strength of $D_0 = 1.2$; only two modes are activated before gain clamping occurs. The electric field intensity of the dominant mode is shown in panel (d).}
    \label{fig:toy}
\end{figure}

To test our numerical approach, we considered the example of a one-dimensional resonator with both ends open and a non-uniform refractive index and pump profile, with frequency of the gain center $k_a=15/L$ where $L$ is the length of the cavity. 
 The refractive index is piecewise constant, with $n = 1.5$ in the left section ($0 \le x \le L/4$) and $n = 3.0$ in the right section ($L/4 < x \le L$). The pump is applied in the region $0 \le x \le L/2$, as shown in Fig.~\ref{fig:toy}. The model is described in~\cite{ge2010steady}.

As expected~\cite{ge2010steady}, only two modes are activated before the gain-clamping. In Fig.~\ref{fig:toy}, the spectrum is represented for the pump strength $D_0=1.2$.

\section{SIPR $\chi^2$ tests}

To quantify the visual differences between the SIPR histograms reported in Fig.~3(a) and Fig.~3(b) of the main text, we performed $\chi^2$ goodness-of-fit tests against two well-known and markedly different distributions: the inverse Gaussian and the gamma--lognormal.

\begin{table}[h!]
\centering
\caption{$\chi^2$ test results for the SIPR distributions shown in Fig.~3 of the main text.}
\label{tab:chi2_SIPR}
\begin{tabular}{l l c l}
\hline
Model & Dataset & $p$-value & Result \\
\hline
Inverse Gaussian & SIPR WD      & 0.9100 & accepted \\
Inverse Gaussian & SIPR Poisson & 0.0008 & rejected \\
Gamma--LogNormal & SIPR WD      & 0.0090 & rejected \\
Gamma--LogNormal & SIPR Poisson & 0.0500 & accepted \\
\hline
\end{tabular}
\end{table}

The SIPR data analyzed are those shown in Fig.~3 of the main text, where panel (a) corresponds to the Poisson fibers length statistics and panel (b) to the Wigner--Dyson fibers length statistics. 
The results are summarized in Table~S1. The tests reveal a clear statistical separation between the two regimes.

\section{Brody Statistics Analysis}

We enrich the Poisson/Wigner–Dyson classification with a continuous
descriptor of statistic in terms of fiber segment lengths. We employ a Brody distribution and
find the corresponding parameter $\beta$ for each network realization, where $\beta = 0$ and $\beta = 1$ correspond
to Poisson and Wigner–Dyson statistics, respectively. We then analyze the spectral
parameters of each configuration of the network as a function of $\beta$. The results show that the
mean values of the spectral features considered (MID, IWD, and SIPR) are similar, as also found
for the other statistics. The results obtained are summarized in Table~\ref{tab:spectral_vs_beta}.

\begin{table}[ht!]
\centering
\caption{Mean values and associated errors for MID, IWD, and SIPR spectral features, categorized by bins of the Brody parameter $\beta$.}
\label{tab:spectral_vs_beta}
\small % Font leggermente ridotto per adattarsi alla colonna
\begin{tabular}{@{}ccccccc@{}}
\toprule
Bins $\beta$ & \multicolumn{2}{c}{MID} & \multicolumn{2}{c}{IWD} & \multicolumn{2}{c}{SIPR} \\
\cmidrule(lr){2-3} \cmidrule(lr){4-5} \cmidrule(lr){6-7}
& Mean & Err & Mean & Err & Mean & Err \\ \midrule
0.00--0.25 & 0.08  & 0.06 & 0.05  & 0.04 & 3.65 & 0.10 \\
0.25--0.50 & 0.02  & 0.06 & 0.05  & 0.04 & 3.99 & 0.12 \\
0.50--0.75 & -0.11 & 0.12 & -0.09 & 0.08 & 3.78 & 0.24 \\
0.75--1.00 & 0.34  & 0.10 & 0.19  & 0.08 & 3.13 & 0.24 \\ \bottomrule
\end{tabular}
\end{table}

%\clearpage
% Bibliography
\bibliography{NLs}

%Manual citation list
%\begin{thebibliography}{1}
%\bibitem{Zhang:14}
%Y.~Zhang, S.~Qiao, L.~Sun, Q.~W. Shi, W.~Huang, %L.~Li, and Z.~Yang,
 % \enquote{Photoinduced active terahertz metamaterials with nanostructured
  %vanadium dioxide film deposited by sol-gel method,} Opt. Express \textbf{22},
  %11070--11078 (2014).
%\end{thebibliography}